\documentclass[%
 reprint,
nofootinbib,
 amsmath,amssymb,
 aps,
]{revtex4-2}

\usepackage[utf8]{inputenc}
\usepackage{xcolor}
\usepackage{subfigure}
\usepackage{epsfig}
\usepackage{amsmath}
\usepackage{makeidx}
\usepackage{amsfonts}
\usepackage{multirow}
\usepackage{dcolumn}
\usepackage{subfig}

\begin{document}

\preprint{APS/123-QED}

\title{Linking Through Time:\\
Memory-Enhanced Community Discovery in Temporal Networks}

\author{Giulio Virginio Clemente}
\email{giulio.clemente@imtlucca.it}
\affiliation{%
 IMT School for Advanced Studies, Piazza San Francesco 19, 55100 Lucca (Italy)
}%

\author{Diego Garlaschelli}
\affiliation{IMT School for Advanced Studies, Piazza S. Francesco 19, 55100 Lucca (Italy)}
\affiliation{Lorentz Institute for Theoretical Physics, Niels Bohrweg 2, 2333 CA Leiden (The Netherlands)}
\affiliation{INdAM-GNAMPA Istituto Nazionale di Alta Matematica (Italy)}%

\date{\today}

\begin{abstract}

Temporal Networks, and more specifically, Markovian Temporal Networks, present a unique challenge regarding the community discovery task. The inherent dynamism of these systems requires an intricate understanding of memory effects and structural heterogeneity, which are often key drivers of network evolution. In this study, we address these aspects by introducing an innovative approach to community detection, centered around a novel modularity function. We focus on demonstrating the improvements our new approach brings to a fundamental aspect of community detection: the detectability threshold problem. We show that by associating memory directly with nodes’ memberships and considering it in the expression of the modularity, the detectability threshold can be lowered with respect to cases where memory is not considered, thereby enhancing the quality of the communities discovered. To validate our approach, we carry out extensive numerical simulations, assessing the effectiveness of our method in a controlled setting. Furthermore, we apply our method to real-world data to underscore its practicality and robustness. This application not only demonstrates the method's effectiveness but also reveals its capacity to indirectly tackle additional challenges, such as determining the optimal time window for aggregating data in dynamic graphs. This illustrates the method's versatility in addressing complex aspects of temporal network analysis.

\end{abstract}

\maketitle


\section{Introduction}
Networks often possess intricate mesostructures that conceal crucial information about the system's functionality. Among these mesostructures, the presence of communities or groups has garnered significant interest within the scientific community. This interest is primarily due to their profound connection with various characterizing aspects of the system, such as the diffusion of processes within it \cite{Diffusion_modularity}, its stability \cite{may1972will, Caruso_block_stability}, and the ability to predict link formation \cite{link_prediction}.

When considering the network's topological structure, communities emerge as clusters of nodes exhibiting a significantly higher link density than expected. Based on this intuition, numerous methodologies have been developed to tackle community detection, ranging from the use of objective functions to quantify the quality of a partition, such as the modularity function \cite{Newman_Girvan}, to approaches based on generative models like the Stochastic Block Model \cite{Decelle2011}, and flow models that leverage network dynamics to cluster nodes based on persistent flow patterns \cite{Rosvall_community}.

Fundamental is the determination of the "expected", a concept at the heart of the primary tool used in this work: the modularity function. This function quantifies the deviation between a given benchmark, which defines what is 'expected,' and the observed number of links within the network. This benchmark, often referred to as a null model, plays a crucial role in obtaining accurate results, a role that we are going to remark in the rest of the paper.

While many of the above-mentioned methodologies have been  well-developed for static networks, they encounter challenges when it comes to generalizing them for time-varying graphs. This difficulty arises primarily due to the additional degree of freedom introduced by time, which needs to be considered. Nevertheless, community detection for temporal networks is a rapidly expanding field, and several studies have endeavored to adapt static methods to the dynamic context \cite{Mucha2010,Barucca2017,Zhang2016}.

When talking about modularity function, it is the absence of appropriate null models that constitutes a significant barrier to generalizing it to a temporal framework. 
In this paper, we fill this gap by demonstrating how the maximum entropy approach \cite{Max_Ent_book} can be utilized to construct various forms of the modularity function and leverage them to account for different aspects of network structure. Specifically, in the subsequent sections, we introduce a novel formulation of the modularity function based on a recent Markovian null model for temporal networks \cite{clemente2023temporal}. This formulation allows simultaneous control over anomalies related to link persistence between nodes and link formation, thus allowing an accurate generalization of the modularity function to time-varying graphs. Additionally, we illustrate how incorporating memory between blocks can generally lead to a dramatic improvements in algorithmic performance.

To demonstrate the advancements achieved by leveraging these properties, we investigate the problem of the Detectability Threshold, which serves as a theoretical limit defining the boundary for the ability of any algorithm to detect the presence of communities \cite{Decelle_detectability}. Initially, we introduce the threshold for the static case, and subsequently, by incorporating the insights presented in \cite{Ghasemian2016}, we compute the threshold while considering the presence of memory between nodes within the same blocks.

We supported our theoretical results with numerical simulations by conducting a series of analysis on synthetic temporal networks, wherein both the number of nodes and memberships remain fixed throughout the network's evolution. Our findings reveal that when the membership of a group of nodes influences both the probability of link formation and the probability of link persistence, algorithm performance consistently improves when considering also information regarding the latter property.

Furthermore, we also conducted empirical analysis by utilizing a real-world dataset of proximity relations in a primary school \cite{SchoolData}. We illustrate how our proposed methodology, can help with the selection of an optimal time window for data aggregation, and can be effectively used for community detection when memory is driven by the nodes' membership.

Our synthetic and empirical analyses aim to underline the importance of considering temporal characteristics and memory in network structures, and to highlight how these considerations can enhance community detection methodology in temporal networks.

\section{Community detection, Null models \\ and Modularity Function}

Consider a graph \(G = (N, E)\), where \(N\) represents a set of nodes and \(E \subseteq N \times N\) denotes the set of edges. A community structure within \(G\) is defined as a partition  $g = \{g_1, g_2, ..., g_q\}$ of \(N\), where each \(g_i \subset N\) is a cluster of nodes such that the density of edges within each \(C_i\) is significantly greater than the density of edges between \(g_i\) and \(N \setminus g_i\). In this context, we focus on non-overlapping communities, implying that each node belongs to exactly one \(g_i\) within \( g \).

Discovering the optimal subdivision for a network with an inherent block structure is a challenging task that has captured the attention of numerous researchers \cite{Fortunato_2009}. To address this issue, Newman and Girvan introduced a quality function in 2004, known as modularity \cite{Newman_Girvan}, to evaluate the effectiveness of network partitions. The original formulation of this function is expressed as follows:

\begin{equation}
Q(g) = \sum_{i,j \in N} \left[ a_{ij} - \langle a_{ij}  \rangle \right] \delta_{g(i) g(j)}
\end{equation}

Where $a_{ij}$ denotes the observed element of the adjacency matrix $A$ for the graph $G$, indicating the presence of a link between node $i$ and node $j$. On the other hand, $\langle a_{ij} \rangle$ represents the expected value of the same link in a null model that assumes the absence of any community structure. The vector $g$ indicates the membership of each node within the community structure.

The modularity function measures the deviation between the observed and expected number of intra-community edges as defined by the null model. Therefore, to identify the optimal partition, our objective is to maximize the modularity function with respect to the assigned node membership, effectively seeking the community structure with the most significant deviation with respect to the specified null model.

When the optimal partition is such that the two quantities are similar, the modularity function approaches zero, indicating a lack of strong community structure. Conversely, when a robust community structure exists, the modularity value tends to approach 1. As already said, it is important to note that the cornerstone of this measure lies in the definition of the null model (a pattern-generating model) that assumes specific properties remain constant while allowing other characteristics to vary stochastically \cite{Gotelli_Graves}. The randomization process aims to recreate certain aspects of the data that would be expected in the absence of any particular mechanisms, such as the presence of a community structure.

It is precisely by leveraging this characteristic that the modularity function accomplishes its objective.

The choice of the null model is subjective and depends on the properties we aim to reproduce in our random network. However, much attention has been given to null models that accurately replicate the degree sequence of undirected and unweighted networks, primarily to study the impact of local inhomogeneity \cite{Newman_null_model}. In this paper, since our main contribution relies on null models that precisely reproduce the degree sequence of a given network, we will exclusively consider null models that achieve this objective.

When constraining the degree sequence, for binary and undirected networks, the widely used specification for $\langle A_{ij} \rangle$ is the well-known Chung-Lu model: $\langle a_{ij} \rangle = \frac{k_i k_j}{2m}$, where $k_i$ and $k_j$ represent the degrees of nodes $i$ and $j$, respectively, and $2m$ is the total number of possible links. However, it is important to note that the Chung-Lu model is an approximation that holds under stringent conditions \cite{Squartini_unbiased}, which many real networks do not satisfy.

An important example of such a deviation is highlighted in the study conducted by Maslov et al. \cite{Maslov}, where they demonstrate that applying the Chung-Lu model to compute the probability of connecting two hubs in a snapshot of the internet network on January 2, 2000 would result in a probability of 43.5. Clearly, this value cannot be interpreted as a probability. Utilizing such a quantity in the modularity function would introduce bias, ultimately impacting the performance of the method.

In this work, we propose a modularity function derived from null models developed through a Maximum Entropy approach \cite{Max_Ent_book}. This approach overcomes the limitations associated with the Chung-Lu model and avoids the aforementioned bias.

\subsection{Maximum Entropy Null Models}

The maximum entropy approach is a methodology that enables the definition of a probability distribution over a set of elements, such as graphs, in a manner that ensures specific expected values are reproduced while all other aspects remain maximally random. This means that if certain characteristics emerge from the model without being explicitly imposed, it is indicative of a genuine correlation between these properties and the constraints utilized in constructing the model.

Applying the maximum entropy formalism to real-world problems involves a two-step process. The first step addresses a graph \textit{G} with certain distinctive properties \textit{C(G)}. In this step, we seek to obtain the most unbiased probability distribution \textit{P(G)} over the set of all possible graphs, ensuring that the expected value of \textit{C(G)} aligns with the imposed constraints. To achieve this distribution, we leverage the Shannon-Gibbs entropy:

\begin{equation}
S(P(G)) = -\sum_{G} P(G)ln(P(G))
\label{eq:entropy}
\end{equation}

Imposing that the probability distribution reproduces on average

\begin{equation}
\langle C \rangle = \sum_G C(G) P(G) = C^*
\label{eq:constraints}
\end{equation}

Where G is an ensemble element, $C^*$ is the measured value of the constraint, and $\langle \cdot \rangle$ indicates the expected value over the ensemble.

It is crucial to emphasize that, as we have defined it, this class of models allows for the imposition of constraints where the expected value, rather than the exact value for each network realization, is reproduced. This has implications not only on computational aspects \cite{Max_Ent_book} but also on the model's ability to replicate metrics not predetermined by constraints \cite{Fluctuating_Networks_2022}.

The constrained maximization problem is approached by employing the method of Lagrange multipliers ($\theta$). The resulting probability distribution has the following functional form \cite{Park_Newman}:

\begin{equation}
P(G|\vec{\theta}) = \frac{e^{-H(G|\vec{\theta})}}{Z(\vec{\theta})}
\label{eq:ERG}
\end{equation}

Where $H(G|\vec{\theta}) = \sum_i \theta_i C_i$ is called graph Hamiltonian, and $Z(\vec{\theta}) = \sum_G e^{-H(G|\vec{\theta})}$ is the normalization constant. It is important to note that the analytical expression for $Z$ is often unknown, making it challenging to obtain an exact expression for $P(G)$. However, in our case, we will consider only situations where solutions have been found.\\
The second step involves determining the numeric values of the parameters $\vec{\theta}$ that reproduce the measured quantities on a real graph. To tackle this problem, we can employ the maximum likelihood principle, which entails setting the parameters to the specific values that maximize the likelihood function $P(G^*|\vec{\theta})$ given the model \cite{likelihood_Garlaschelli_loffredo}.

The first maximum entropy model we introduce is the Binary Configuration Model \cite{Park_Newman}, which serves as the counterpart to the Chung-Lu model constructed using the maximum entropy formalism. As previously mentioned, this represents the primary model employed in the classical definition of modularity.
The Binary Configuration Model is obtained by imposing constraints on the degree sequence, leading to the following form of the Hamiltonian:

\begin{equation}
H(\mathbf{G}|\vec{\theta})\equiv\sum_{i=1}^N \theta_i k_i
\label{eq:Hstatic}
\end{equation}
Where $k_i$ is the degree of node \textit{i}.
Hence for this specific problem, we have that
\begin{equation}
\langle a_{ij}  \rangle \equiv \frac{x_i x_j}{1+x_i x_j}.
\label{eq:pstatic}
\end{equation}

where $x_i\equiv e^{-\theta_i}$ for each node $i$ \cite{likelihood_Garlaschelli_loffredo,Max_Ent_book}.

In the subsequent sections, we will utilize the maximum entropy formalism to employ specific null models for defining new modularity expressions and characterizing their properties.

\subsection{Detectability Threshold In Static Graphs}

In order to explore the impact of temporal information on community detection, we investigate the behavior of different modularity functions in relation to the detectability threshold problem. This phenomenon has garnered significant attention from researchers in recent years \cite{Reichardt_detect_thres, Detectability_threshold}. It has been demonstrated that there exists a fundamental limitation to the ability of algorithms to detect community structures in networks, characterized by a phase transition known as the detectability threshold (DT) \cite{Detectability_threshold}.

To introduce this issue, we consider networks generated using the well-known Stochastic Block Model (SBM) \cite{Holland1983}. The SBM is a generative model that produces networks with a block structure, making it a suitable framework for studying the performances of community detection algorithms in a controlled environment. By analyzing networks generated by the SBM, we can gain insights into the behavior of community detection algorithms and investigate how temporal information can potentially enhance their performance.

Given a set of nodes \textit{N} and \textit{q} groups denoted by $g = \{g_1, g_2, ..., g_q\}$, the SBM in its basic formulation defines a network structure in which unweighted edges are placed between node \textit{i} and node \textit{j} with probability $p_{g_i g_j}$ depending on the groups which they belong. Hence the probability of observing a graph \textit{G} is:

\begin{equation}
    P(G|g) = \prod_{(i j) \in N} p_{g_i g_j}^{a_{i j}} \left( 1-p_{g_i g_j}     \right)^{1-a_{i j}}
    \label{SBM}
\end{equation}

Where $a_{ij}$ is the element of the adjacency matrix A corresponding the graph $G$.\\
In the following, we will refer to a particular class of block model defined by clusters having the same dimension: Planted Partition Model, specifically all the theoretical results we show are related to a Planted Partition model with two communities that will be characterized by a probability of connection between nodes of different clusters $p_{out}$ and within nodes of the same cluster $p_{in}$.\\
For this specific case, it has been shown rigorously \cite{Mossel_threshold} that given a sparse graph \textit{G}, i.e. with $p_{ij}=O(\frac{1}{N})$, with two communities, an average degree \textit{k}, an average number of connections between groups $k_{out} = \frac{N}{2} p_{out}$, and within groups $k_{in} = \frac{N}{2}  p_{in}$, it exists a value $k_{in} - k_{out} = \Delta^*$ such that, above this threshold (\textbf{DT}), no algorithm can capture the community structure better than by chance:

\begin{equation}
    \Delta^* = 2 \sqrt{k}
    \label{eq:dt_static}
\end{equation}

This expression of the threshold represents a result that will be used subsequently as a reference to demonstrate the benefit of using multiple realizations of the same network, rather than the single network on which the previous expression was obtained.

In the following, to measure the quality of the partition computed by our models, we use the the Adjusted Rand Index (ARI) \cite{rand_score} defined in the following:

{\small
\begin{equation}
    ARI(g_{t}, g_{e}) = \frac{\sum_{ij} \binom{n_{ij}}{2} - \left[ \sum_i \binom{a_i}{2} \sum_j \binom{b_j}{2} \right] / \binom{N}{2}} 
    {\frac{1}{2} \left[ \sum_i \binom{a_i}{2} + \sum_j \binom{b_j}{2} \right] - \left[ \binom{a_i}{2} \sum_j \binom{b_j}{2} \right] / \binom{N}{2}}
\end{equation}
}

Where, index \textit{i} goes over all the sets in $g_t$, while index \textit{j} goes over all sets in $g_e$.\\
$n_{ij}$ defines the number of elements in common between the sets in $g_t(i)$ and $g_e(j)$,  $a_i = \sum_j n_{ij}$ is the sum of the elements in common between set \textit{i} and all sets in \textit{j}, and $b_j \sum_i n_{ij}$ is the sum of the elements in common between the set \textit{j} and all the possible sets \textit{j}. Hence, the value of ARI varies between 0, when none of the elements is well classified, and 1, when there is a perfect matching between the real membership and the one inferred.
In this way we can compare the communities obtained by the modularity maximization with the one given as ground-truth.

\begin{figure}[!h]
\centering
\includegraphics[width=5.9cm]{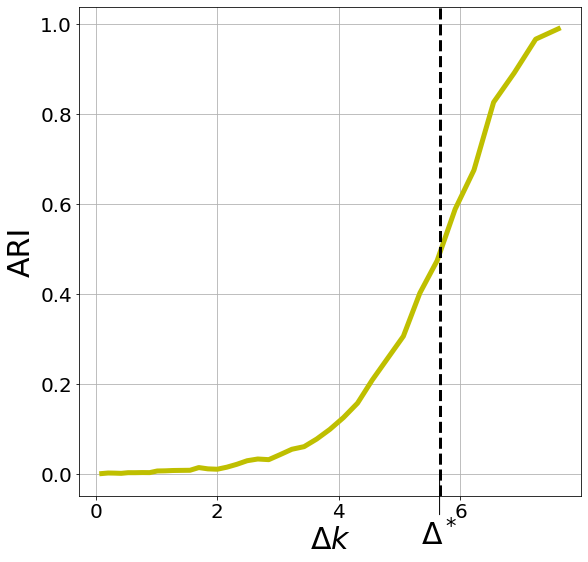}
\caption{ In figure are shown the performances using the modularity defined in eq. (\ref{eq:static_modularity}) on a static network with k = 8 and 200 nodes. Each point is averaged over 30 iteration.}
\label{fig:static_detect}
\end{figure}

As a first experiment, we use the modularity function defined with the null model in eq. (\ref{eq:pstatic}), denoted as:

\begin{equation}
Q(g) = \sum_{i,j} \left[ a_{ij} - \frac{x_i x_j}{1+x_i x_j} \right] \delta_{g(i) g(j)}
\label{eq:static_modularity}
\end{equation}

we can examine how the community detection algorithm behaves when varying the relationship between $p_{in}$ and $p_{out}$. Since the expected degree is the same for each node, a suitable null model for our case involves a single parameter, in contrast to the binary configuration model (CM) that requires \textit{N} parameters. However, the choice of the null model does not significantly affect the final results, so we employ the binary configuration model for illustrative purposes. Figure (\ref{fig:static_detect}) displays the transition between a regime where the modularity function-based algorithm successfully detects the community structure and a regime where it fails to do so.

The displayed plot results from the examination of graphs, each composed of two communities, in which the relationship between $p_{in}$ and $p_{out}$ varies as previously described. These artificially generated graphs maintain a constant average node degree. The plot substantiates the theoretical forecast that a distinct transition point, denoted as $\Delta = \Delta^*$, exists. The experiments confirms the existence of a point that demarcates the boundary between regions where community structure is detectable and where it remains undetectable.

\section{\centering Modularity for Temporal Networks with Stationary Data Generating Process}

The transition from static to temporal graphs brings an added dimension that can significantly impact tasks like community detection. The network's evolution over time, as noted in \cite{Survey_temporal_community}, can be driven by alterations in  the community structure.

To deal with these evolving systems, it is necessary to acknowledge that temporal networks can be represented through various methods. Thus, selecting an approach that best aligns with our proposed algorithm is crucial.
For our study, we face the community detection problem working with a snapshot representation, namely a sequence of \textit{T} static sparse graphs, each denoted as $G_i$. In our context, every graph in the sequence maintains a constant number of nodes, \textit{N}, and is characterized as unweighted and undirected. We refer to this series as a 'graph trajectory', which is denoted as follows:

\begin{equation}
{ \mathcal{G}}\equiv \{\mathbf{G}_1,\dots,\mathbf{G}_T\}
\end{equation}

As said, communities in temporal networks can experience a variety of transformations \cite{Palla_2007}. Our focus, however, is on persistent communities that actively shape the network's evolution. Though this represents a very simple case of study, it is worth investigating because, as we'll show, it helps to demostrate how the right null model can greatly boost community detection accuracy, and, in this specific case, to show how memory can be enlightening in community discovery.

\begin{figure}[!h]
    \centering  
    \includegraphics[width=8.5cm]{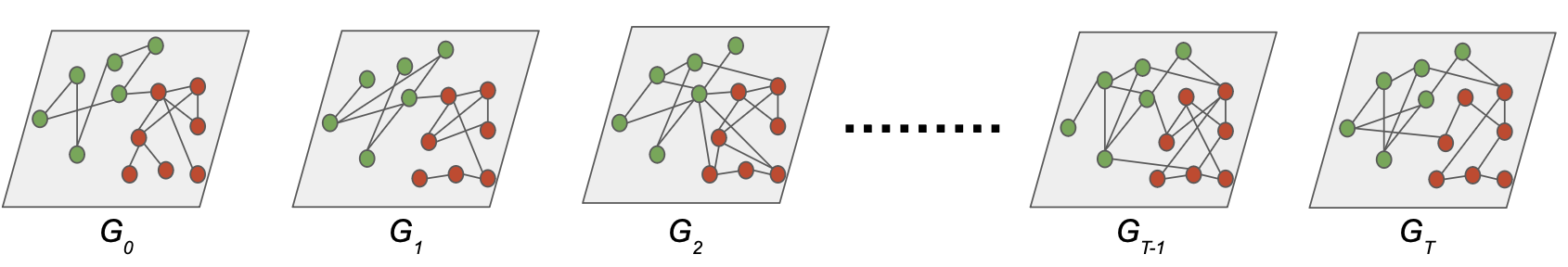}
    \caption{Schematic Representation of Temporal Networks with Fixed Communities. Each node in the network is assigned a color to indicate its membership that remains fixed throughout the evolution of the network. }
    \label{fig:Ex_TN}
\end{figure}

Our study postulates that networks evolve with stable community memberships and parameters, as illustrated in Fig. \ref{fig:Ex_TN}. This implies that communities are assumed to maintain consistency across the graph trajectory, generated by a stationary process. We particularly explore two different data generating processes: initially, a time-varying graph with no correlation between snapshots, subsequently evolving into a temporal network with inherent memory.

\subsection{\centering Temporal Networks with independent observations}

The first Data Generating Process (DGP) that we analyze generates a graph trajectory characterized by a sequence of independent and sparse snapshots, each of them built according to a  Planted Partition Model, the very same model used for the static case. In this way the average connectivity for each block in each snapshot is equal to \textit{k}, implying that the average connectivity in time, $\sum_t \frac{1}{T} k(t) = \overline{k} = k$.
Hence, the model so defined has the following expression, representing the probability of observing a given graph trajectory $\mathcal{G}$:

\begin{equation}
    P(\mathcal{G}|g) = \prod_{t=1}^T \prod_{(i j)} p_{g_i g_j}^{a_{g_i g_j}(t)} \left( 1-p_{g_i g_j}     \right)^{1-a_{g_i g_j} (t)}
    \label{DSBM_stationary}
\end{equation}

We refer to this model as a Dynamic Stochastic Block Model (DSBM).

To address community detection within this framework, we employ a modularity function that utilizes as a null model a naive temporal generalization of the Binary Configuration Model. This model is constructed by substituting the degree sequence of a single snapshot with its temporal average as a constraint. Consequently, the related Hamiltonian is expressed as follows:

\begin{equation}
H(\mathcal{G}|\vec{\alpha})\equiv  \sum_{i=1}^N \alpha_i  \sum_{t=1}^T \frac{k_i(t) }{T}
\label{eq:Hcmdynamic}
\end{equation}

Formally, the resulting expression for the expected value of the link existence between two nodes is the same as the non-temporal binary configuration model, but this time it refers to its average over time:

\begin{equation}
    \langle \overline{a_{ij}}  \rangle = \frac{x_i x_j}{1+x_i x_j}
    \label{eq:a_ij_dynamic_cm}
\end{equation}

Where, in this case, $x_i  = e^{-\alpha_i/T} $.\\
Therefore, the related modularity can be defined as:

\begin{equation}
Q(g) = \sum_{i,j} \left[ \  \overline{a_{ij}} - \langle \overline{a_{ij}}  \rangle \ \right] \delta_{g(i) g(j)}
\label{no_memory_modularity}
\end{equation}

With $\langle \overline{a_{ij}}  \rangle$ as in eq. (\ref{eq:a_ij_dynamic_cm}) and $\overline{a_{ij}} = \frac{1}{T} \sum_T a_{ij}(t) $.

Intuitively, when the community structure does not change in time, as the case presented in this section, using eq. (\ref{no_memory_modularity}) allows to filter out the spurious links by observing several realizations of the same network, hence directly putting more evidence on the hidden block structure.

However, by employing this modularity function, we implicitly assume that each snapshot represents an independent realization of the same network. This approach overlooks the temporal order in the graph's trajectory, thereby discarding relevant information when it is present.

\subsection{\centering Community detection in Markovian \\ Time-Varying Graphs}\label{sec:Time_Varying_Graphs}

We now aim to introduce a refined version of the modularity function, which incorporates node features associated with the nodes' capacity to maintain links.
These ideas are derived from the findings presented in \cite{clemente2023temporal}, where the authors demonstrate the importance of capturing node-level memory and how a model incorporating this aspect may be crucial for analyzing real-world systems.
Based on this consideration, here ahead, we want to demonstrate how, by leveraging these features, we can improve the performances of the modularity function on the community detection task.
To achieve the aforementioned objectives, we consider a DGP with a Markovian structure, which allows us to exert control over the memory embedded in the system. 

The DGP employed can be seen as a Dynamic Stochastic Block Model that differs from others of its kind found in the literature \cite{Barucca2017, Ghasemian2016,dall2020}. Indeed, contrary to other DGPs where memory is an external parameter, independent of the network's characteristics such as degree distribution, in our model, we incorporate a memory effect. This effect is embodied in a Markovian evolution of the system, governed by node-specific features that simultaneously control link formation and persistence, interdependently.

Therefore, to introduce the model, we begin by identifying the features to be encoded in the maximum entropy model, which will serve as the foundation for both the DGP and the definition of modularity. In this context, we focus on what is termed the 'persisting degree', the portion of the degree that is maintained between one snapshot and the next:

\begin{equation}
h_{i}(t,1)\equiv\sum_{j\ne i}a_{ij}(t)a_{ij}(t+1) = \sum_{j\ne i}b_{ij}(t,1).
\label{eq:hdef}
\end{equation}

We will refer to the quantity $a_{ij}(t)a_{ij}(t+1)$ as the one-lagged \textit{persisting link}.\\

Hence, by imposing this as further constraints at the case described by the Hamiltonian in eq. (\ref{eq:Hcmdynamic}), we generate the following Hamiltonian:

\begin{equation}
 H(\mathcal{G}|\vec{\alpha},\vec{\beta})\equiv  \sum_{i=1}^N  \left[ \alpha_i  \sum_{t=1}^T \frac{k_i(t) }{T} + \beta_i  \sum_{t=1}^T \frac{h_i(t) }{T}     \right]
\label{eq:Hkh}
\end{equation}

By solving the related problem \cite{clemente2023temporal}, we can obtain the analytical expression for the expected value of the average link and the average persisting-link:
\begin{equation}
p_{ij}=\langle \overline{a_{ij}(t)}\rangle
\label{marginal}
\end{equation}
and 
\begin{equation}
q_{ij} = \langle \overline{a_{ij}(t)a_{ij}(t+1)}\rangle
\label{persisting}
\end{equation}
Both the expressions as function of $\vec{\alpha}$ and $\vec{\beta}$ (see the supplementary information for the details). \\
Where $p_{ij}$ represents the marginal probability of having a link between node \textit{i} and node \textit{j} at time \textit{t}, and

\begin{equation}
    P[a_{ij}(t+1) = 1 \ \& \ a_{ij}(t)=1   ] =  q_{ij}    
\end{equation}

is the probability that two nodes have a connection at time t and $t+1$.

Once obtained $p_{ij}$ and $q_{ij}$,  we can compute the following stochastic matrix:

\begin{equation}
P_{ij}(q_{ij},p_{ij}) = 
\begin{bmatrix}
\frac{q_{ij}}{p_{ij}} & \frac{p_{ij}-q_{ij}}{p_{ij}} \\

\frac{p_{ij}-q_{ij}}{1-p_{ij}}  &  \frac{1-2p_{ij}+q_{ij}} {1-p_{ij}}
\end{bmatrix}
\label{Stochastic_matrix}
\end{equation}

Which defines the transition matrix associated with the evolution of each link.

This stochastic matrix essentially defines the system's evolution, ensuring that the expected values for the persisting degree and the degree precisely match those constrained in eq. (\ref{eq:Hkh}).

Operationally, our version of the Dynamic Stochastic Block Model begins with the generation of an initial network snapshot using a static Planted Partition Model. Subsequently, we allow the link presence to evolve according to a Stochastic Matrix in eq. (\ref{Stochastic_matrix}), creating the full trajectory.

Similar to the other cases, in our experiments we generate graph trajectories by specifying our DGP with two distinct probabilities: \(p_{out}\) and \(p_{in}\) for connection, and \(q_{out}\) and \(q_{in}\) for preserving links. These probabilities automatically induce an average degree, \textit{k}, and an average persisting degree, \textit{h}, respectively. Consequently, the membership influences both the link density within blocks and the persisting link density. All probabilities are set to ensure the network's sparsity is maintained, being of the order \(O(\frac{1}{N})\), which is a crucial condition for defining the expression for the Detectability Threshold.

Once the graph sequence $\mathcal{G}$ is generated, we can obtain two sequences of matrices: 

\begin{equation}
\mathcal{A} = \left(  A(1), A(2) ... A(T) \right)
\end{equation}

And

\begin{equation}
\mathcal{B} = \left(  B(1), B(2) ... B(T) \right)
\end{equation}

Where $A(t) = \left( a_{ij}(t) \right)$ and $B(t) = \left( a_{ij}(t+1)a_{ij}(t) \right)$.\\

From which we can define a new matrix 

\begin{equation}
C(t) = A(t) + B(t)    
\label{matrix_c}
\end{equation}

Such that each element of $C_{ij}(t) = a_{ij}(t) + b_{ij}(t)$.

By computing the average in time of A(t) and B(t), we obtain the matrix $\overline{\mathcal{C}}$, whose elements are used to define a new modularity in the following way:
\begin{equation}
Q(g) =  \sum_{ij \in N} \left[  \overline{c_{ij}(t)}  - \langle \overline{c_{ij}} \rangle  \right] \delta(g_i,g_j)
\label{temporal_modularity_2}
\end{equation}
 Where $\langle \overline{c_{ij}} \rangle = p_{ij} + q_{ij}$.
 
Through this approach, we account for two contributions that are, in principle, interdependent yet capable of providing distinct insights into the DGP.

\section{Detectability Threshold for Stationary Time-Varying Graphs}

This section illustrates how the DT varies based on the assumptions about the DGP and the model applied for data analysis. As previously stated, the DGP remains constant over time, indicating that our data are samples from a stationary process. Under this premise, it is intuitive that an increase in the number of samples, i.e., the length of the trajectory, reduces the influence of noise on the observed block structures. This reduction occurs as noise is averaged out over more samples, consequently enhancing the DT, and leading to improved outcomes.

Nevertheless, this intuition is not the whole story. Indeed, as we are going to show, using the right features and a suitable model to capture those features, can positively change the value of the DT and the results of the algorithm.
Building upon findings in the literature, we demonstrate how the implementation of the above-mentioned Maximum Entropy null model for temporal networks alters the expression for the DT and the regions where the algorithm works well increase.

\subsection{Detectability Threshold for non-correlated Time-Varying Graphs}

We now recall the expression for the DT in a first and well known case, where we consider a DGP in which every snapshot is sparse and independent, as in eq. (\ref{DSBM_stationary}).

It is known that under the condition of sparsity and independence, an expression for the DT has been obtained by \cite{Ghasemian2016} for $T \rightarrow \infty$ and conjectured for finite T, a conjecture that has then been demonstrated in \cite{dall2020}.
The focus of the computation relies on the sparsity of each snapshot that implies a locally tree-like structure for the graph trajectory $\mathcal{G}$. Hence the authors in \cite{Ghasemian2016,dall2020}, interpreted the community detection task in temporal networks as a label recovery problem, showing that, for our particular case in which communities remain stable in time, community detection is feasible only if:

\begin{equation}
    \Delta_{nm}^* = 2 \sqrt{\frac{\overline{k}}{T}}
\end{equation}

Where we define $\Delta_{nm}\footnote{nm stands for no-memory.} = \overline{k_{in}} - \overline{k_{out}}$.
For this kind of temporal network, all the information related to the block structure is contained in the degree sequence, so the threshold has an expression that is a direct function of this quantity. 
Interestingly, as anticipated, the threshold is also a function of T and is such that as T increases $\Delta_{nm}^*$ goes to zero, so, if T is big enough, we will always spot the block structure. 

Given this framework, here we can employ a modularity function that uses only information on the degree sequence to spot the community structure.

\subsection{Detectability Threshold for Markovian Time-Varying Graphs}\label{sec:DT_Memory}

Transitioning to Markovian temporal networks introduces correlations between snapshots, effectively embedding intrinsic memory. In our model, this memory is assimilated by enabling link persistence, a feature previously noted for its undesirable effects on community detection task performance \cite{Barucca2017}.

Introducing snapshot correlations complicates the observed matrix sequence \(\mathcal{A}\), with links influenced more by memory than indicative of community structure. 

Furthermore, the lack of independence between consecutive snapshots leads to the loss of prerequisites for computing the DT, as described in \cite{Ghasemian2016,dall2020}. However, it is worth noting that within our DGP, we can precisely determine the necessary waiting period to consider two consecutive matrices uncorrelated, thereby restoring the conditions required to calculate the threshold.

This can be done by acknowledging that each link in our graph evolves according to the stochastic matrix defined in eq.(\ref{Stochastic_matrix}). This matrix is characterized by two eigenvalues: the first, by definition, is equal to 1, and the second is given by:

\begin{equation}
    \lambda_{ij}^{(2)} = \frac{q_{ij}-p_{ij}^2}{p_{ij}-p_{ij}^2}
    \label{eigenvalue_2}
\end{equation}

This eigenvalue is directly linked to the system's memory. Indeed, by calculating \( \lambda_{ij}^{(2)} \), as detailed in \cite{clemente2023temporal}, we can determine the correlation length and the rate at which the correlation diminishes between two consecutive snapshots. Consequently, we obtain \(\tau^*\), which is defined as:

\begin{equation}
    \tau_{ij}^* = \frac{1}{ln\left(\frac{1}{\lambda_{ij}^{(2)}}\right)}
    \label{tau_star}
\end{equation}

\begin{figure}[!h]
\centering
\includegraphics[width=8cm]{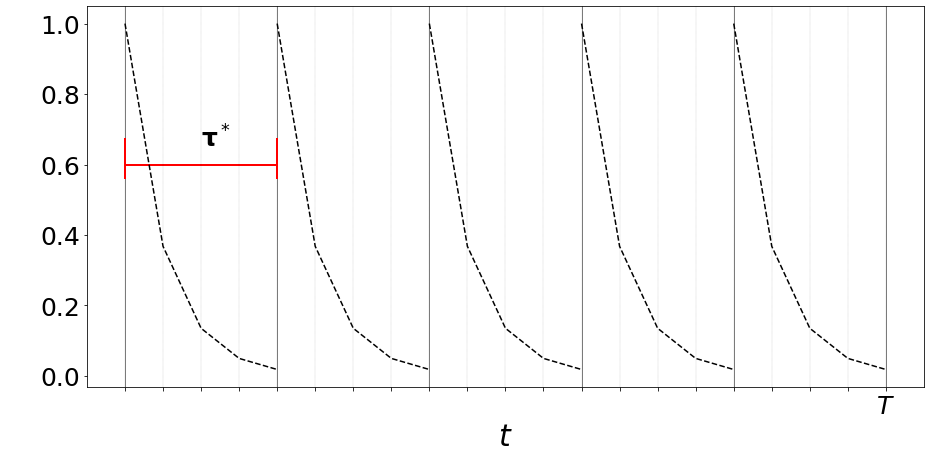}
\caption{In the figure, a schematic representation of the behavior for the memory in time for system evolving according to eq. (\ref{Stochastic_matrix}). $\tau^*$ represents the relaxation time.}
\label{behavior_of_memory}
\end{figure}

As schematically represented in Fig. (\ref{behavior_of_memory}), waiting for a duration equal to the maximum of all \(\tau_{ij}^*\), denoted as \(\tau^*\), between one snapshot and the next ensures that the snapshots can be considered independent.

At this point, by discounting the entries of the matrix $\overline{\mathcal{A}}$ by the maximum correlation length $\tau^*$, we reproduce the condition for the calculation of the DT for finite T,
resulting in the following expression for the DT:

\begin{equation}
    \Delta_{M,nm}^* = 2 \sqrt{\frac{\tau^*}{T} \overline{k}}
    \label{eq:DT_M_NM}
\end{equation}

Where M stands for "Markovian".

Moving ahead, we can state that in scenarios where the process exhibits memory, relying only on the degree sequence to identify communities results in the neglect of crucial information that can provide insights into the block structure. In the following, we will demonstrate how incorporating the persisting degree enhances the algorithm's performance and changes the expression for DT.

In this regard, our innovative modularity function, as defined in eq. (\ref{temporal_modularity_2}), integrates information from two distinct matrices, $\overline{\mathcal{B}}$ and $\overline{\mathcal{A}}$. This approach allows us to present how the method proposed by Ghasemian et al. \cite{Ghasemian2016} to compute the DT can be re-adapted to our context, thereby implying a new and better threshold.

We have previously demonstrated that, by adjusting \(\overline{\mathcal{A}}\) with \(\tau^*\), its elements can be interpreted as those realized by an equivalent DSBM with independent snapshots \cite{dall2020, Barucca2017}. To extend this rationale to the entire matrix \(\overline{\mathcal{C}}\), it is necessary to justify how the same discounting factor can be applied to the matrix \(\mathcal{B}\). This is achieved by explicitly calculating the autocorrelation between elements at time \(t\) and \(t + 1\) for matrix \(\mathcal{B}\).

As detailed in the Supplementary Information, by adjusting for a time \(\tau^*\), even in this scenario, the elements of matrix \(\mathcal{B}\) can be considered independent. This adjustment allows for the application of the previously discussed methodology in calculating the DT. Thus, when considering the matrix resulting from the sum of \(\mathcal{A}\) and \(\mathcal{B}\), which remains sparse, we re-establish the conditions necessary for the analytical computation of the DT:

\begin{equation}
    \Delta_{M,m}^* = 2 \sqrt{\frac{\tau^*}{T} \overline{c}}
\end{equation}

Where $c = \overline{k} + \overline{h}$, and $\overline{h} < \overline{k}$ by definition.
This result shows that, if the block membership guides the link persistence, this property can be used to improve the quality of the discovered community structure.


\section{Numerical Simulations}\label{sec:Numerical_Simulations}
To validate the analytical results presented in the previous sections, we have conducted a series of numerical simulations. These simulations demonstrate the consistency between the predicted and observed thresholds. Additionally, we characterize the algorithm's performance across various combinations of parameters, showing how our novel modularity function works better in different scenarios.

To achieve this, we utilize the Data Generating Processes previously introduced for temporal networks, along with their associated modularity functions. For all experiments, we optimize the modularity function using the algorithm detailed in \cite{Algorithm}, adapting the expression for the null model specific to each case. We generate the trajectories by varying the values of \(\Delta k\) and \(\Delta h\), as defined in:

\begin{equation}
    \Delta k = k_{in} - k_{out} \quad \textit{and} \quad \Delta h = h_{in} - h_{out}
    \label{eq:dk_dh}
\end{equation}

with 

\begin{eqnarray}
    k_{in} = \frac{N}{2}p_{in},  \quad  h_{in} = \frac{N}{2}q_{in} \\ 
    k_{out} = \frac{N}{2}p_{out}, \quad  h_{out} = \frac{N}{2}q_{out}
\end{eqnarray}

Here, \(N\) denotes the number of nodes within a single snapshot. In general, each trajectory is generated to maintain both the average degree and the average persisting degree across the entire temporal network, when those are specified, or otherwise only the average degree. 

In this setting, \(\Delta k\) signifies the disparity between the average (temporal) number of links formed within the communities and those outside, whereas \(\Delta h\) denotes the difference in the persistence of links within these communities compared to external links.

In all the experiments, we have two communities whose elements are known and play the role of ground-truth communities. 

The initial series of experiments involves generating graph trajectories with independent observations, as in eq. (\ref{DSBM_stationary}), by varying \(k_{in}\) and \(k_{out}\), thus altering the value of \(\Delta k\) while maintaining a constant average degree \(k\). By specifying the number of time steps for each trajectory, we consequently establish the threshold \(\Delta_{nm}^*\).

We generate 50 trajectories for each \(\Delta K\) value and various \(T\) values, then plot the average performance across these 50 instances for the Community Detection task.

\begin{figure}[!h]
\centering
\includegraphics[width=5.9cm]{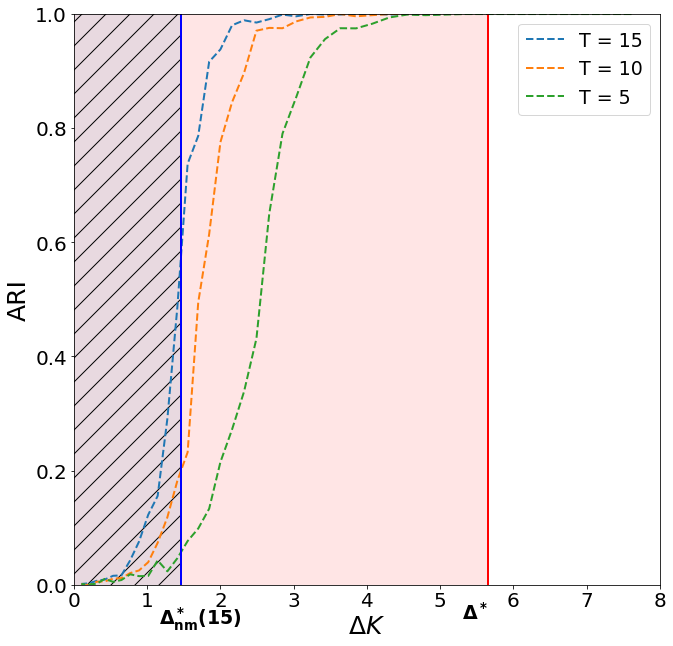}
\caption{The figure represents three different profiles for the performances of the community detection task on graph trajectories of different lengths. The vertical line blu represents the DT for the temporal network with T=15, while the red vertical line is for a single static snapshot.}
\label{DGP_no_memory}
\end{figure}

In Figure (\ref{DGP_no_memory}), are represented the results obtained using the modularity function described in eq. (\ref{no_memory_modularity}). The red vertical line corresponds to the static DT eq.(\ref{eq:dt_static}) as computed on the first snapshot of the trajectory (see Fig. \ref{fig:static_detect}). It is interesting to note how even with few snapshots, the gain of performance quality is evident. As expected, in all the cases, the DT predicted and the ones computed numerically are the same.

In our next analysis, we explore graph trajectories that are generated through a Markovian data-generating process, as detailed in eq. (\ref{Stochastic_matrix}). Similar to our previous experiments, we generate 50 trajectories, each for different values of \(T\) and \(\Delta K\). However, in this updated scenario, we incorporate a memory effect by keeping the values of \(q_{in}\) and \(q_{out}\) constant, but allowing \(p_{in}\) and \(p_{out}\) to vary. This approach, as discussed in Section (\ref{sec:DT_Memory}), is expected to significantly influence the performance of the modularity, previously defined in eq. (\ref{no_memory_modularity}).

\begin{figure}[!h]
\centering
\includegraphics[width=5.9cm]{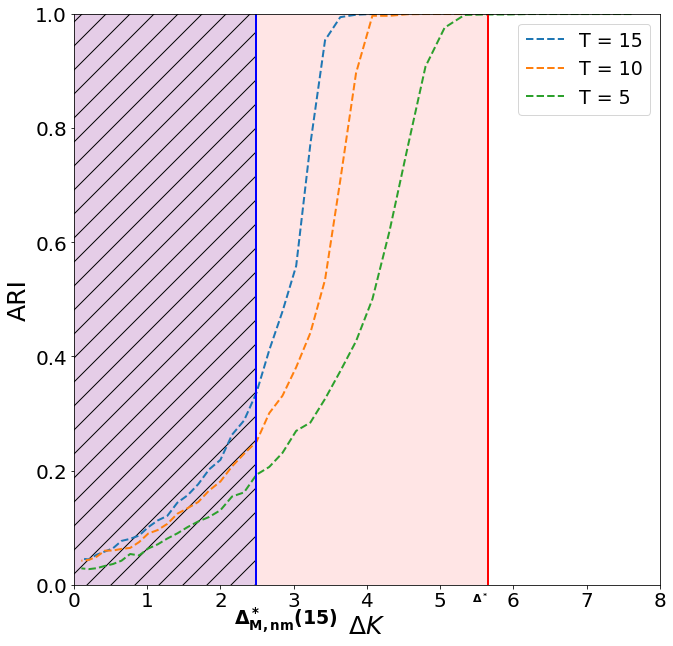}
\caption{In figure is displayed the average ARI values for different values of $\Delta K$ and T. The vertical lines represent the computed thresholds for the static case (in red) and the dynamic case with T=15 (in blue).}
\label{DGP_memory}
\end{figure}

 Figure (\ref{DGP_memory}) illustrates the adverse effects of memory on the quality of the detected communities. This diminishes the benefits of using additional snapshots compared to the static case, thereby confirming the observations in eq. (\ref{eq:DT_M_NM}). Notably, the curve patterns in Fig. (\ref{DGP_memory}) differ from those observed in Fig. (\ref{DGP_no_memory}). Intrinsically, the value of \(\tau^*\), which reflects the memory level in our process, is influenced not only by \(q_{in}\) and \(q_{out}\) but also by \(\Delta K\), as expressed in eq. (\ref{eigenvalue_2}). This dependency impacts the memory and so, it alters the curve shapes. 

Having demonstrated how memory can impact on the ability to discover communities in temporal networks, we now explore how it can be leveraged to recover some of the lost information about community structure. We proceed to the most comprehensive scenario addressed in this study, generating graph trajectories while varying \(\Delta k\) and \(\Delta h\) for different values of \(T\). It's important to note that the average value of \(\overline{h}\) should remain lower than \(\overline{k}\) during trajectory generation in order to create realistic trajectories, which, of course, it influences the value of \(\tau^*\). In fact, as we adjust these two parameters, while maintaining the average values of \(\overline{k}\) and \(\overline{h}\) constant, the resultant changes in \(\tau^*\) significantly affect both the curve shapes and the value of the DT as already mentioned and showen in Fig. (\ref{DGP_memory}).

To realize this task we use the Dynamic Stochastic Block model described in Sec. (\ref{sec:Time_Varying_Graphs}).

In the experiments we choose all the parameters such that we keep the threshold $\Delta_m*$ fixed, but varying the relationships between $k_{in}, k_{out},h_{in}$ and $h_{out}$, so to explore the algorithmic performances and see whether the threshold predicted analytically is validated numerically. 

Given the process used to generate the trajectories, we ensure that the evolution does not change the relation between $p_{in}$ and $p_{out}$, facilitating the direct comparison between the modularity built on the model with memory and the one built on the memory-less one.

\begin{figure*}[!htb] 
\begin{minipage}{0.32\textwidth}
  \centering
  \includegraphics[width=\linewidth]{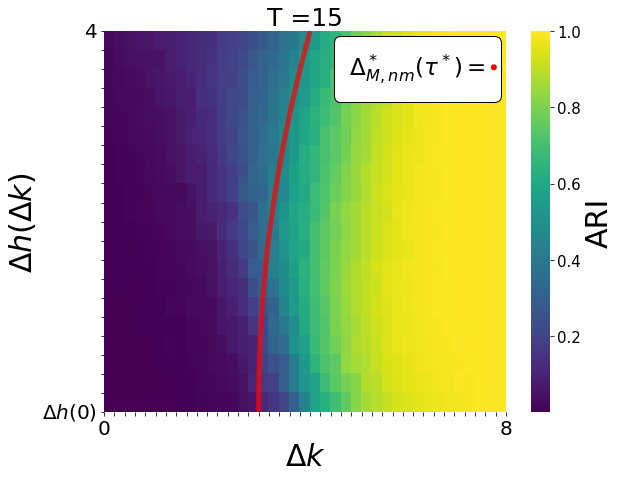}
  \caption*{(a)}
  \label{fig:no_memory_15}
\end{minipage}\hfill
\begin{minipage}{0.32\textwidth}
  \centering
  \includegraphics[width=\linewidth]{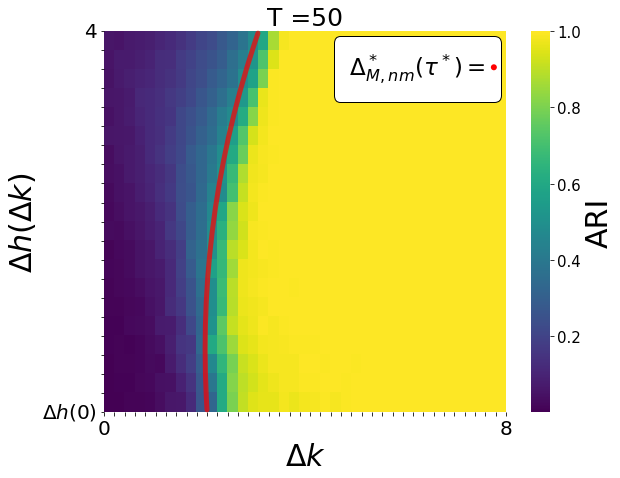}
  \caption*{(b)}
  \label{fig:no_memory_50}
\end{minipage}\hfill
\begin{minipage}{0.32\textwidth}
  \centering
    \includegraphics[width=\linewidth]{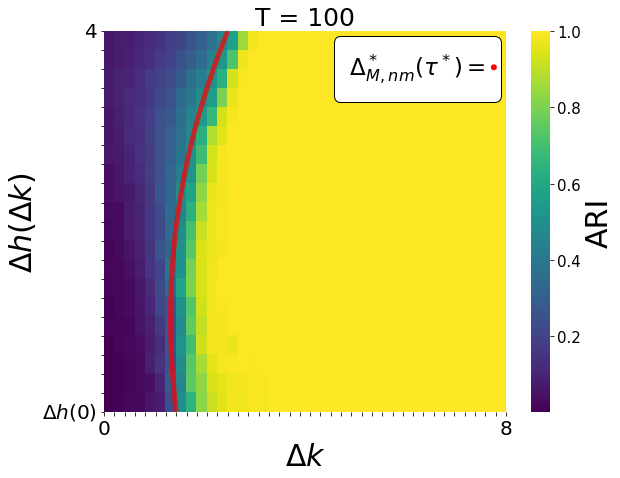}
  \caption*{(c)}
  \label{fig:no_memory_100}
\end{minipage}

\begin{minipage}{0.32\textwidth}
  \centering
  \includegraphics[width=\linewidth]{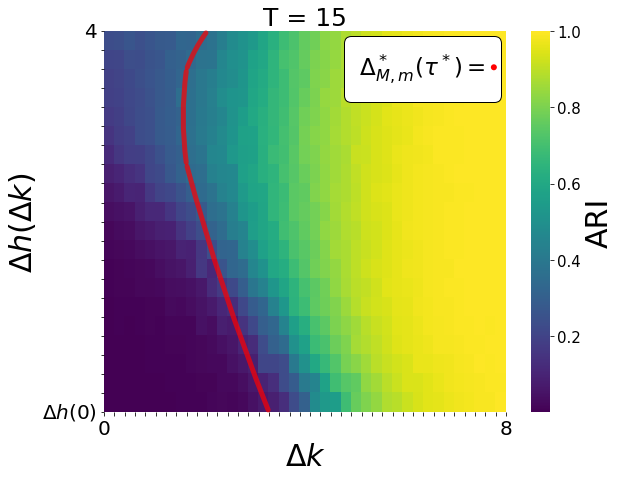}
  \caption*{(d)}
  \label{fig:memory_15}
\end{minipage}\hfill
\begin{minipage}{0.32\textwidth}
  \centering
  \includegraphics[width=\linewidth]{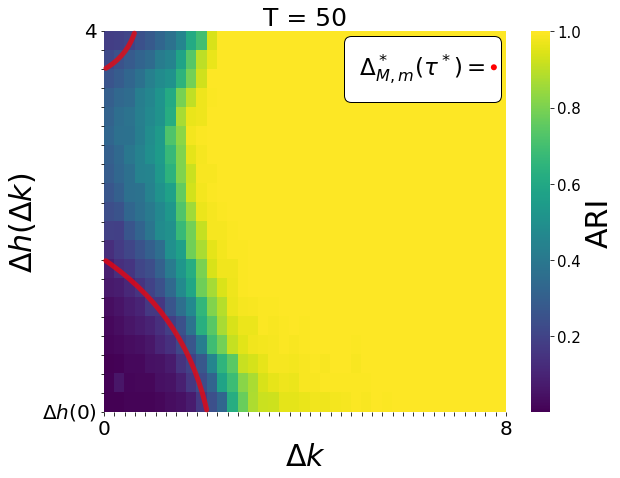}
  \caption*{(e)}
  \label{fig:memory_50}
\end{minipage}\hfill
\begin{minipage}{0.32\textwidth}
  \centering
  \includegraphics[width=\linewidth]{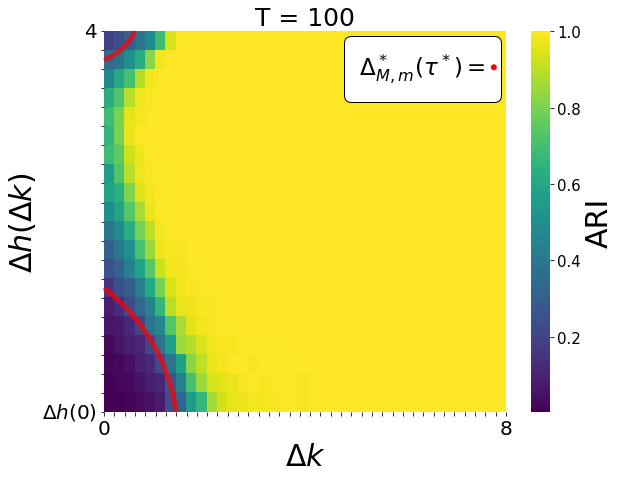}
  \caption*{(g)}
  \label{fig:memory_100}
\end{minipage}

\caption{In these figures, we present three heatmaps for each of the following modularity functions: the \textbf{no-memory modularity function} (Figures a, b, and c at the top) and the \textbf{memory modularity function} (Figures d, e, and f at the bottom). These heatmaps illustrate the performance on three Markovian networks of increasing lengths (T = 15, T = 50, and T = 100, respectively). In each plot, the analytical DT is represented by a red line.}
\label{fig:combined_threshold}
\end{figure*}

The experimental outcomes, as depicted in Fig. (\ref{fig:combined_threshold}), align with expectations. The no-memory approach fails to recognize the varying propensity of nodes within different groups to preserve links, inadvertently discarding valuable information.

As the number of samples generated by the same DGP increases, the accuracy of parameter estimation improves. Consequently, a greater number of snapshots leads to more precise inference of the community structure.

It is evident that \(\Delta_{nm}^* > \Delta_{m}^*\), while there are zones where, specifically for \(\Delta h(0)\), the distinction between modularity incorporating the persisting degree and that without it becomes negligible. This is attributed to the neglectable disparity at this point between the persisting links within and outside communities. 

However, as illustrated in Fig. (\ref{fig:combined_threshold}), for all other scenarios, incorporating this additional information consistently enhances the quality of the detected communities, showing how our novel modularity performs better.

\section{Proximity Networks}\label{sec:Real_Data}

We now proceed to examine a real-world application, aiming to test the implementation of the previously defined modularities on a dataset where ground-truth community information is accessible.

The dataset provides information about proximity relations between students and teachers in a primary school \cite{SchoolData}. Electronic proximity detectors worn by the participants were used to collect the data, capturing the presence of other nearby detectors at 20-second intervals. The data was collected over a period of two days, specifically from October 1st to October 2nd, 2009. For simplicity, in this analysis, we focus only on the data from the first day.
Based on this data, we aim to construct a temporal network, which will be characterized by a series of snapshots created by aggregating data over a specific time window. This implies that within each window, two nodes will be connected if they have come into contact at least once during that time interval.
In our study, we leverage verified information about existing communities. The underlying hypothesis is that the network dynamics are primarily driven by the community memberships of the nodes. The aim is to investigate whether, by examining these dynamics, it's possible to discern the node memberships. It's important to note that these dynamics are far from trivial, as even during class hours, students change classrooms and have the freedom to move around.

Working with snapshot representations of real temporal networks requires to take into account how the choice of time windows for aggregating information can impact the quality of results, as noted by Krings et al \cite{Krings} and Clauset et al \cite{periodicity_net}. 
Before suggesting a solution to this challenge for our specific scenario, we begin analyzing the real network by initially adopting a 10-minute time aggregation window. This initial aggregation will be used for preliminary analyses, which will lay the foundation for developing a strategy to determine the optimal time window for our task.

The proposed approach for community identification is based on the assumption that the process that generated the data is stationary. Therefore, it becomes crucial to identify temporal segments for which it makes sense to consider the network as generated by the same process. To solve this task and identify potential breakpoints in the system's evolution, we apply the approach introduced in \cite{clemente2023temporal}, by using the model that takes into account memory at node level as the reference model.

We have to emphasize that in choosing periods using this model, we are implicitly assuming that the observed stationary period is one during which the memory among links remains relatively stable.

We assume that the network is generated using a Dynamic Stochastic Block Model using the ground truth communities, and we evaluate the performances of our algorithm. We then observe the behavior of the algorithm and the quality of the communities found in the various identified segments.


\begin{figure}[h!]
\centering
\includegraphics[height=5.0cm, width=5.5cm]{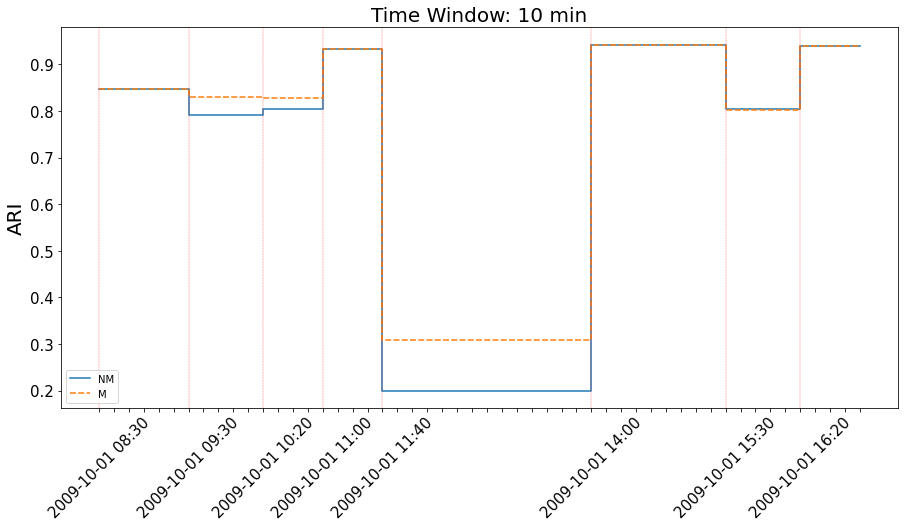}
\caption{The figure displays the ARI values for 10 minutes temporal windows at various identified breakpoints. Additionally, the values for different types of null models are indicated.}
\label{fig: ARI_real_net}
\end{figure}

In Fig. (\ref{fig: ARI_real_net}) are presented the performances of the algorithm when executed over a 10-minute time window. We identify several changing points that reflect diverse dynamic behaviors. In this particular scenario, these dynamics are understood to result from the students' class memberships and their movements across various classrooms in the school. It is noteworthy that our approach finds that influence of class-based membership diminishes significantly during periods in the playground.
This result allows us to deduce that the supposed data generating process, the Dynamic Stochastic Block Model with which the trajectory has formed, can be observed during class hours. However, during playground time, the decrease in the quality of the observed communities compared to the ground truth leads us to believe that the assumed data generating process may be incorrect during these hours, and that the dynamics within that interval are not well described by this model.

We now take a further step in our investigation by examining how the quality of the communities identified changes with different time windows. Our objective is to determine if there is a natural scale where the effects of membership on link density and persistent link density are more evident. To this end, we distinguish between class hours, where an higher similarity between the discovered communities and the ground truth was noted, and playground hours. Our algorithm is applied across various time windows.
For each time window, we identify distinct stationary segments. In each segment, we assess the Adjusted Rand Index between the communities detected using our modularities and the actual communities. Additionally, we characterize each stationary segment using \(\Delta k\) and \(\Delta h\).

In this context, \(\Delta k\) is defined as the difference between the average (temporal) number of links formed within the ground-truth communities and those formed outside, while \(\Delta h\) represents the difference in the number of links that persist within the ground truth communities compared to those outside, hence the terms in eq. (\ref{eq:dk_dh}) becomes:

\begin{eqnarray}
    k_{in} = \sum_{ij} \overline{a}_{ij}^* \delta(g_i^*,g_j^*),  \quad  h_{in} = \sum_{ij} \overline{b}_{ij}^* \delta(g_i^*,g_j^*) \\ 
    k_{out} = \sum_{ij} \overline{a}_{ij}^* \textit{I}_{(g_i^* \neq g_j^*)}, \quad  h_{out} = \sum_{ij} \overline{b}_{ij}^* \textit{I}_{(g_i^* \neq g_j^*)}
\end{eqnarray}

with, \textit{I} as indicator function

\begin{equation}
\textit{I}_{(g_i \neq g_j)} = 
\begin{cases} 
1 & \text{if } g_i \neq g_j, \\
0 & \text{otherwise}.
\end{cases}    
\end{equation}

\begin{figure*}[h! tbp]
    \begin{minipage}{0.48\textwidth}
        \includegraphics[height=6cm, width=8cm]{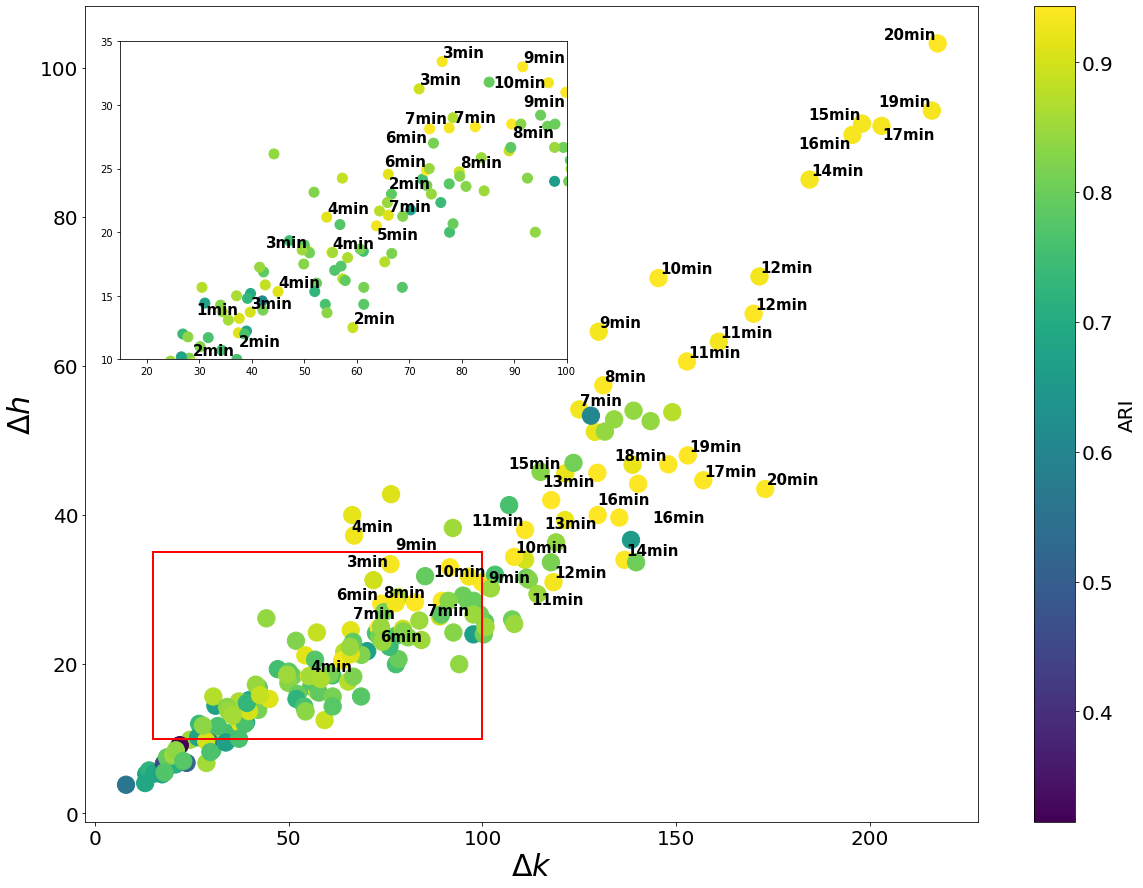}
        \caption*{(a)}
        \label{fig: dk_dh_class1}
    \end{minipage}
    \begin{minipage}{0.48\textwidth}
        \includegraphics[height=6cm, width=8cm]{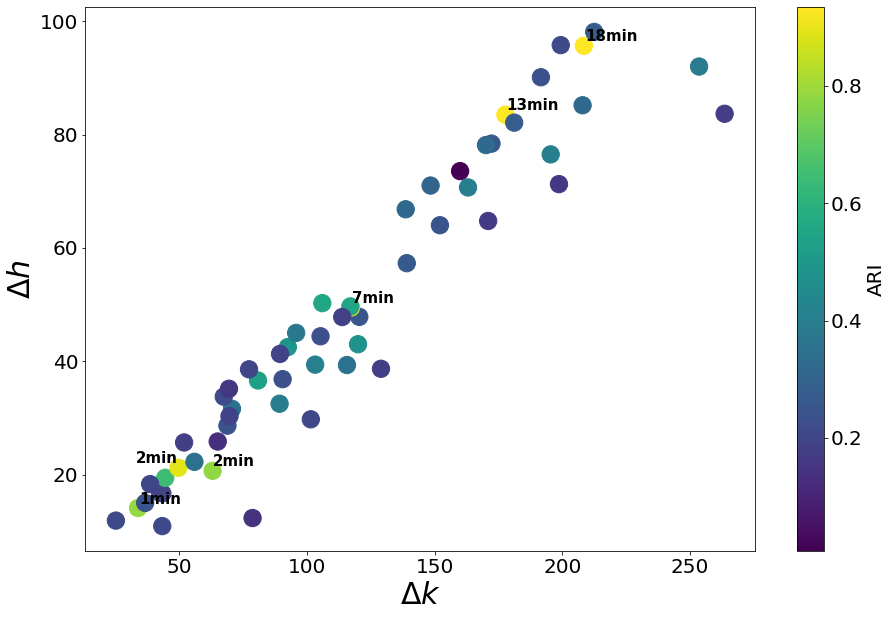}
        \caption*{(b)}
        \label{fig: dk_dh_class2}
    \end{minipage}
    \caption{Figure displays the performance of community detection on a real network. Each point corresponds to a stationary segment identified using structural break detection. To each point, a value of \(\Delta k\) and \(\Delta h\) is attributed. The color of the points indicates the ARI value, calculated by comparing the communities detected with the modularity with memory against the ground truths. In plot (a), the values during class hours are shown, while plot (b) depicts the values during playground.The label associated with each point indicates the size of the temporal window used to create the temporal network associated with that segment. Only labels corresponding to ARI values greater than 0.8 are displayed.}
    \label{fig:dk_dh_class3}
\end{figure*}

Figure (\ref{fig:dk_dh_class3}) presents the outcomes of this analysis. It reveals that during class hours, an increase in both \(\Delta k\) and \(\Delta h\) is associated with a corresponding rise in the ARI. However, during playground hours, as previously observed in Fig. (\ref{fig: ARI_real_net}), this correlation appears to be absent.
It's important to emphasize that in this case, $\Delta k$ and $\Delta h$ do not assume the same meaning as in cases involving only two partitions.  Here, what we observe is the cumulative sum computed across all possible clusters, which amounts is greater then two in our dataset.
Hence, it might be possible to observe an increase in the difference between links due to the contribution of one of the communities for which there was a reference and not in an homogeneous way for all the clusters.

So our results show that for the non-playground part the performances, regardless of the aggregation window, seem to follow the expectations for a scenario where the evolution is genuinely driven by membership. As the temporal window increases, there is an increase in \(\Delta k\) and \(\Delta h\) for certain segments, which the model can leverage to improve the quality of the communities. During class hours, memory and community structures appear to be related.

The situation differs during playground hours. In this latter case, it's probable that other effects are mixed with the existing ones, rendering the presumed Dynamic Stochastic Block Model with the class membership less suitable for accurately describing the system.

To strengthen the relationship between memory and the quality of the detected communities, we propose an additional analysis. We examine if there is a relationship between the presence of memory and the quality of the observed communities: the hypothesis is that if the system's evolution follows the membership of the elements, then there may exist a natural temporal scale at which both the quality of the communities, measured with the ARI, and the memory are at their peak. This natural time scale would result in the preferred time window for aggregating data in such a way that the communities become more distinct.
We measure memory with normalized autocovariance. The normalization is done in such a way to allow us to control for the effect of degree. In general, the average normalized autocovariance will be higher if the degree is larger. By dividing by the degree of each node, we discount this effect.

Hence, as a metric to this end, we employ the average autocovariance, defined as follows:

\begin{eqnarray}
\overline{\textbf{A}}(W,\tau=1) = \frac{1}{N}\sum_i^N \frac{A_i(\tau=1)}{k_i}
    \label{eq:norm_mod_c}
\end{eqnarray}

Here, the average is computed over all nodes \textit{N} in the network, and

\begin{equation}
     A_i(\tau)= \frac{1}{T}\sum_{t=1}^T\big[ h_i(t,\tau)-\langle h_i(t,\tau)\rangle_{nm}\big]
\end{equation}

represents the autocovariance specific to node \textit{i}. In this context, $\langle h_i(t,\tau)\rangle_{nm}$ denotes the expected node degree persistence according to a network randomization model that precludes memory.
To accomplish this, we utilize a null model specifically designed to not contemplate memory effects, as defined in eq. (\ref{eq:a_ij_dynamic_cm}). This approach retains any genuine temporal correlations between snapshots, enabling a more accurate investigation of memory effects.

Having identified class hours as the temporal segments where dynamics are driven by membership, we focus on these to study the behavior of normalized autocovariance. Our strategy involves varying the time window within a range of 1 to 20 minutes:

\begin{equation}
W \in \{1', 2', 3', \ldots, 20'\}
\end{equation}

We identify all the stationary segments that fall within class hours using the same approach as before, corresponding to the different time windows. For each of these, we calculate the normalized autocovariance for $\tau = 1$. We  use the average of the normalized autocovariance across all identified stationary segments as a measure of memory. Thus, defining \(\textit{brks}\) as the set of segments where we can assume the memory to be constant, the memory associated with each time window will be defined as:

\begin{equation}
\mu_A(W) = \frac{1}{|\textit{brks}|} \sum_{s \in \textit{brks}} \overline{\textbf{A}_s}(W,\tau=1)    
\end{equation}

Hence, once for each segment, is computed the expected value of the persisting degree relative to a memoryless model (\(\langle h_i(t,\tau) \rangle_{\text{nm}}\)), we can compute $\mu_A$.

At this stage, we need to establish a criterion for selecting the optimal time window, which will be utilized for evaluating the efficacy of our community detection modularity.

We define the "best window" \( W_{best} \) as the index \(  W \) for which \( \mu_{A}(W) \) is maximum:

\begin{equation}
    W_{best} = \arg \max \mu_{A}(W)
\end{equation}

\begin{figure}[!h]
\centering
\includegraphics[height=3cm, width=8cm]{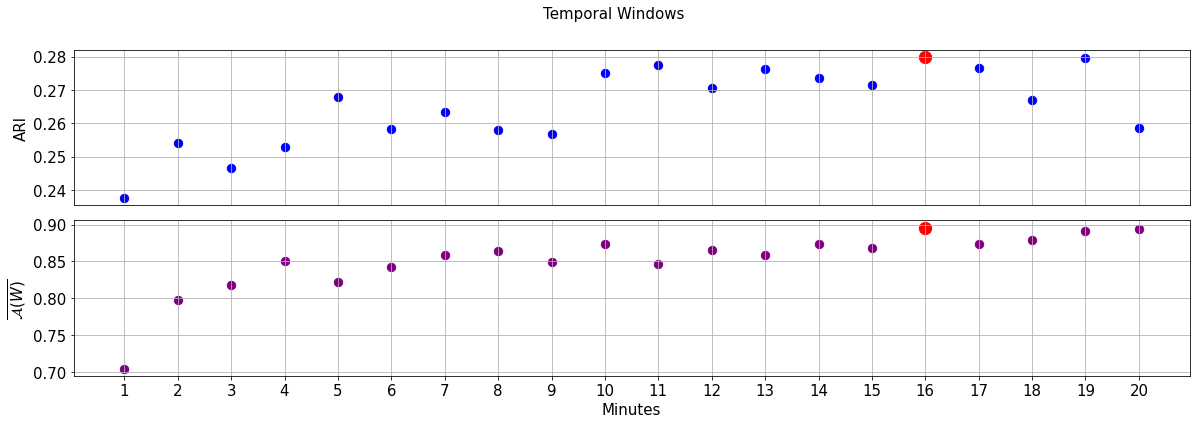}
\caption{The figure represents the normalized autocovariance measurement across three time segments for different time window.}
\label{fig:autocovariance}
\end{figure}

in Fig. (\ref{fig:autocovariance}) is presented the measurement of the normalized autocovariance as function of the time window and the average ARI computed over the segments in class hours. Each point on the plot is designed for a non-overlapping time window. The red points represent the maximum value taken between all the windows. 
Between these two quantities we measure the Pearson Correlation, funding a strong and significative correlation with value \textbf{0.73}, demostrating a relationship between the presence of memory and the quality of the communities found.

In our data, $W_{best} = 16 \ \textit{minutes}$, which also corresponds to the maximum on the ARI.

In support of this assessment, we have included controlled experiments reported in the supplementary information (\ref{SI:Results_Memory}) where the system evolves exactly following these the hypothesis tested above.

\section{Discussion}

Effective community detection in temporal networks necessitates the development of new, appropriate null models \cite{Bassett_et_al}. Such models may provide a basis for comparison and enable the identification of dynamic structures in networks enhancing the quality of the discovered communities.
In this work, we make a contribution in this direction by demonstrating how the use of the right null model can lead to the discovery of hidden structures with higher precision.

We show the potential of modularity-based methods in community detection by utilizing Maximum Entropy null models to tailor their formulation to the specific challenges encountered. We propose a new modularity function that acknowledges the persistence of connections among nodes within the same community.  By focusing on the issue of the detectability threshold, our findings reveal that selecting an appropriate null model can substantially elevate the quality of the identified communities. This highlights the critical role of accounting for memory effects, when present, and explains how incorporating memory changes the detectability threshold for community identification.

In real-world applications, we demonstrate the importance of considering memory in community discovery, and how this feature can assist in indirectly related challenges,  such as determining the ideal time window for data aggregation to develop a time-varying graph. We propose a data-driven methodology to identify the most suitable time window that reveals a more distinct delineation of community structures.

The results indicate that if memory is present but not adequately incorporated into the model, it can deteriorate the quality of the identified communities. However, if properly integrated into the algorithm used for community detection, it can become a fundamental advantage.

In numerous real-world systems, the influence of community membership on memory is a critical factor that cannot be ignored. Under these circumstances, a function that integrates memory effects, such as the one we propose in this study, is essential. It holds the potential to uncover new patterns that were previously hidden. By acknowledging and incorporating memory, our approach opens the door to revealing deeper insights into the dynamics of these systems.

\section{acknowledgements}
This work is supported by the European Union - NextGenerationEU - National Recovery and Resilience Plan (Piano Nazionale di Ripresa e Resilienza, PNRR), project `SoBigData.it - Strengthening the Italian RI for Social Mining and Big Data Analytics' - Grant IR0000013 (n. 3264, 28/12/2021) (\url{https://pnrr.sobigdata.it/}).
D.G. acknowledges support from the Dutch Econophysics Foundation (Stichting Econophysics, Leiden, the Netherlands) and from the project NetRes - `Network analysis of economic and financial resilience', Italian DM n. 289, 25-03-2021 (PRO3 Scuole), CUP D67G22000130001 (\url{https://netres.imtlucca.it}).

\appendix
\section{Supplementary Information}

\subsection{Explicit Expressions for $p_{ij}$ and $q_{ij}$}

The expressions for \(p_{ij}\) and \(q_{ij}\) are derived from the solution to the problem defined by Hamiltonian in eq. (\ref{eq:Hkh}), here recalled:

\begin{equation}
 H(\mathcal{G}|\vec{\alpha},\vec{\beta})\equiv  \sum_{i=1}^N  \left[ \alpha_i  \sum_{t=1}^T \frac{k_i(t) }{T} + \beta_i  \sum_{t=1}^T \frac{h_i(t) }{T}     \right]
\end{equation}

In \cite{clemente2023temporal} it is provided an exact formulation for these quantities. Specifically, through appropriate reparametrization, the model outlined by the Hamiltonian can be precisely mapped to a combination of non-interacting one-dimensional Ising models. These models are then solvable analytically, with their solutions expressible in terms of a constant connection probability.

\begin{eqnarray}
 &p_{ij}= \left(\frac{x_{i}x_{j} y_{ij} -1}{2\sqrt{4x_{i}x_{j} + (x_{i}x_{j} y_{ij} - 1)^2}} \right)\left( \frac{(\lambda_{ij}^+)^T - (\lambda_{ij}^-)^T}{(\lambda_{ij}^+)^T + (\lambda_{ij}^-)^T} \right)  +\frac{1}{2}& \nonumber \\
 \\
&q_{ij}(\tau)=p_{ij}^2+p_{ij}(1-p_{ij})\left(  \frac{(\lambda_{ij}^-)^\tau (\lambda_{ij}^+)^{T-\tau}  + (\lambda_{ij}^+)^\tau (\lambda_{ij}^-)^{T-\tau}    }{(\lambda_{ij}^+)^T + (\lambda_{ij}^-)^T} \right)& \nonumber\\
\end{eqnarray}
With 
\begin{eqnarray}
\lambda_{ij}^{\pm} = e^{J_{ij}} \cosh B_{ij}\pm\sqrt{e^{2J_{ij}}\sinh^2 B_{ij}+e^{-2J_{ij}}}
\end{eqnarray}
and 
\begin{eqnarray}
B_{ij}&\equiv& \frac{1}{2}\left[ \ln(x_{i})+ \ln(x_{j})+\ln(y_{i}) + \ln(y_{i})\right]\\
J_{ij}&\equiv& \frac{1}{4}\left[ \ln(y_{i}) + \ln(y_{j}) \right]
\end{eqnarray}

With, $x_{i} = e^{-\alpha_i/T}$, while $y_{i} = e^{-\beta_i/T} $.

Based on these terms, we can determine the probability of observing or not observing a link at time \(t + \tau\) given its presence or absence at time \(t\). This concept is then translated in the stochastic matrix defined in eq. (\ref{Stochastic_matrix}).

\subsection{Auto-covariance for the matrix B}

Our main argument for the computation of the DT in the case in which modularity using the persisting degree is employed exploits a heuristic, based on the assumption that after $\tau^*$ define in equation (\ref{tau_star}) all the matrices being in the sequence $\mathcal{B}$ and $\mathcal{A}$ can be considered independent.

Using the argument in \cite{covariance_B}, we compute:

\begin{eqnarray}
Cov(B_{ij}(t),B_{ij}(t+\tau)) = \\ \nonumber
Cov\left( a_{ij}(t)a_{ij}(t+\tau),a_{ij} (t+\tau)a_{ij}(t+2\tau) \right)   
\end{eqnarray}

Where, by conducting the calculations, we have the result as follows:

\begin{widetext}
\begin{eqnarray}
   &Cov(B_{ij}(t),B_{ij}(t+\tau)) = \nonumber&\\
   &E[a_{ij}(t) ]^2 \big[ Cov(a_{ij}(t),a_{ij}(t+\tau))^2 
   + Var(a_{ij}(t)) + Cov(a_{ij}(t),a_{ij}(t + 2\tau))  \big] + \nonumber&\\
   &+ Cov(a_{ij}(t + \tau))^2 + Var(a_{ij}(t)) Cov(a_{ij}(t),a_{ij}(t + 2\tau))&
\end{eqnarray} 
\end{widetext}

Where we use the fact that $Cov(a_{ij}(t),a_{ij}(t+\tau) = Cov(a_{ij}(t+\tau),a_{ij}(t + 2\tau))$

That, after substituting the values of the covariances as computed in \cite{clemente2023temporal} and the expected values, becomes:

\begin{widetext}
\begin{eqnarray}
    &Cov(B_{ij}(t),B_{ij}(t+\tau)) =& \nonumber\\
    &= p_{ij}^2 \big[ p_{ij}(1-p_{ij}) + 2g_{ij}(\tau) + g_{ij}(2\tau)  \big] + g(\tau)^2 + p_{ij}(1-p_{ij})g_{ij}(2\tau) =& \nonumber \\
    &= p_{ij}^3(1-p_{ij}) + p_{ij}^2g_{ij}(2\tau) + 2p_{ij}^2g(\tau) + g_{ij}(\tau)^2 + p_{ij}(1-p_{ij})g_{ij}(2\tau)&
    \label{covariance_b}
\end{eqnarray}
\end{widetext}

Where $g(\tau)$ is a function of $\tau$, defined in \cite{clemente2023temporal}, whose value decays with $\tau$:
\begin{equation}
g(\tau) = p_{ij}(1-p_{ij}) (\mu)^{\tau}
\end{equation}
with $\mu < 1$.
Considering \ref{covariance_b}, we note that after $\tau^*$, and considering the term $p_{ij}^3(1-p_{ij})$ small enough, it is equal to 0.

\subsection{Generating Community-Preserving Graph Trajectories via Markov Chains}\label{sec:generate_tn}

Operationally, we begin by defining a network generated according to a stochastic block model characterized by an internal cluster connection probability, \( p_{in} \), and an external connection probability, \( p_{out} \). At this point, two different stochastic matrices are used to define the evolution of connections in this initial network. These matrices are characterized by \( p_{in} \) and \( q_{in} \) for internal cluster connections, and \( p_{out} \) and \( q_{out} \) for external connections. By varying the ratio between the densities of link and link persistence, we are able to study the behavior of our methodology.

Trajectories are generated in a way that preserves both the network's average total degree \(\bar{k}^*\) and the average total persistent degree \(\bar{h}^*\). 

The concept revolves around altering the ratio between the average contribution from internal cluster connections and that from external links. This goal is achieved by adjusting \(p_{in}\) and \(p_{out}\) as well as \(q_{in}\) and \(q_{out}\). To maintain \(\bar{k}^*\) and \(\bar{h}^*\) as mentioned, we use the following formulas.

\begin{eqnarray}
    p_{in} \frac{N}{n_{groups}} + p_{out} \frac{N (n_{groups} - 1)}{n_{groups}} = \overline{k}^* \\
    q_{in} \frac{N}{n_{groups}} + q_{out} \frac{N (n_{groups} - 1)}{n_{groups}} = \overline{h}^*
\end{eqnarray}

Where \( n_{groups} \) represents the number of clusters, \( N \) is the total number of nodes in the network. 

Hence by choosing $p_{out}$ and $q_{out}$ as function of $p_{in}$ and $q_{in}$:

\begin{eqnarray}
    p_{out} = \epsilon p_{in}  \\
    q_{out} = \eta q_{in} 
\end{eqnarray}

Where \( \eta \) and \( \epsilon \) are two parameters that define the internal and external link/persisting link density within the clusters.

We can vary the difference between the internal cluster and the external one, by working on these two parameters.
Using this further conditions we obtain the following formulas:

\begin{eqnarray}
\left\{
\begin{array}{l}
p_{out} = \epsilon p_{in} \\
p_{in} = \frac{n_{groups} \overline{k}^*} {\left [  1 + \epsilon( n_{groups} - 1 ) \right] N }  
\end{array}
\right.
\label{eq:pin_pout}
\end{eqnarray}

\begin{eqnarray}
\left\{
\begin{array}{l}
q_{out} = \eta q_{in} \\
q_{in} = \frac{n_{groups} \overline{h}^*} {\left [  1 + \eta( n_{groups} - 1 ) \right] N }  
\end{array}
\right.
\label{eq:qin_qout}
\end{eqnarray}

By using equations (\ref{eq:pin_pout}) and (\ref{eq:qin_qout}), we ensure that the expected value on the trajectory of \(\left < \bar{h} \right > = \bar{h}^*\) and \(\left < \bar{k} \right > = \bar{k}^*\).

By changing \(\eta\) and \(\epsilon\), we can increase or decrease the disparity between the internal and external link density.
Specifically these parameters can assume values in the following intervals:

\begin{equation}
    \epsilon \in \left\{ kv \mid k \in \mathbb{N}, \, nv \leq 1 \ \& \ nv > 0\right\}
\end{equation}

\begin{equation}
    \eta \in \left\{ r\epsilon \mid r \in (0,1)\right\}
\end{equation}

In this way, the network's evolution will preserve the condition whereby the persistent degree within the group and outside the group will always be lower, respectively, than the internal degree to the group and the external one.

In the numerical simulations presented in Section \ref{sec:Numerical_Simulations}, we focused on a scenario where \(n_{groups} = 2\).

\subsection{A Schematic Presentation of the Algorithm}

Here we explain the steps followed to optimize the modularity function.
The algorithm aims to maximize the modularity function by changing the labels associated with each element of the network.
We exploit the method proposed by Clauset and Moore in \cite{Algorithm}, by readapting it to use the Maximum Entropy null models.
In this sense we rewrite the modularity function in a general form, so that the function for which we want to find the maximum is:

\begin{equation}
    Q(g) = \sum_{i,j} \left[  C_{ij} - P_{ij}^{(null)} \right] \delta_{g(i) g(j)}
\end{equation}

Where depending on the features considered, $C_{ij}$ is:
\begin{equation}
C_{ij} = 
\begin{cases} 
a_{ij}  \quad  &\text{Static Case}\\
\overline{a_{ij}} \quad &\text{Dynamic Case No Memory}  \\
\overline{a_{ij}} + \overline{b_{ij}} \quad &\text{Dynamic Case With Memory}  
\end{cases}
\label{eq:C_ij}
\end{equation}

While $P_{ij}^{(null)}$ assumes the form of the null models presented in the main text.
The implementation proposed in \cite{Algorithm} works by defining a matrix $\Delta Q$, whose elements are indicated with $\Delta Q_{vw}$ (defined below) for each couple of communities. The size of this matrix is initially assumed to be equal to the number of nodes, i.e. at each node is initially attached a unique label. The size is then reduced according to the merging procedure, so to reduce the number of communities.
Hence, for our application, we define the initial value of $\Delta Q_{vw}$ as:

\begin{equation}
\Delta Q_{vw} =  C_{vw} - P_{vw}^{(null)} \quad \text{for} \quad j,i \in N
\end{equation}

Then is selected the largest $\Delta Q_{vw}$, and if it is positive, communities \textit{v} and \textit{w} are merged, the value of Q is incremented of $\Delta Q_{vw}$ and the matrix $\Delta Q$ is updated according to the following rule:

\begin{equation}
\Delta Q_{vk}^{new} =  \Delta Q_{vk} + \Delta Q_{wk}
\end{equation}

The procedure is repeated until it is no more convenient to merge communities.

\subsection{Analyzing the Evolution of Detected Communities and the Role of Memory}\label{SI:Results_Memory}

The purpose of this section is to present additional analyses that support the findings from the main sections. In the first subsection, we conduct numerical simulations to demonstrate how selecting the appropriate time window for data aggregation correlates with the observed memory in the data, as measured by the normalized autocovariance. This selection is relevant in enhancing the algorithm's ability in detecting communities. In the second subsection, we compare the performance of modularity with memory against modularity without memory across various time windows, illustrating the distinct advantages of each approach.

\subsubsection{Numerical Simulations time windows}\label{subsec:Numerical_Simulations_Memory}

We carried out a series of numerical experiments to verify whether, in line with the findings from Section \ref{sec:Real_Data}, the point at which the normalized autocovariance peaks also corresponds to a more discernible community structure as detected by the memory-exploiting modularity.
In our simulations, we begin with a time-varying graph generated as described in Section \ref{sec:generate_tn}, featuring a structure of 5 communities. From this graph, we create additional structures by disaggregating and aggregating snapshots. The disaggregation process involves dividing each snapshot from the original time-varying graph into a number of sub-snapshots. Each sub-snapshot $G_{sub}$ comprises links randomly selected from the original snapshot, ensuring that the sum of all sub-graphs reconstructs the original, hence:

\begin{equation}
\bigcup_{t=1}^{n} G_{sub}(t) = G^{original}
\end{equation}

Where the operator $\bigcup$ operates in such a way that if a link is present in both snapshot $t_1$ and snapshot $t_2$, it is counted only once in the aggregated graph.\\
Temporal order is maintained; for example, if the first snapshot is divided into \(n\) graphs, these will occupy the same position as the original snapshot in the complete time-varying graph.
The aggregation process functions by summing multiple snapshots to form a new snapshot, which will contain all the links from the combined structures. The procedure ensures that there will be non-overlapping snapshots.
Through this approach, we establish a controlled setting with a natural scale at which memory is incorporated into the generative process. This enables us to numerically check whether this scale corresponds to the point where the normalized autocovariance is higher and where the modularity, with known ground-truth communities, performs optimally.

\begin{figure}[h!]
\includegraphics[height=3.9cm, width=6.5cm]{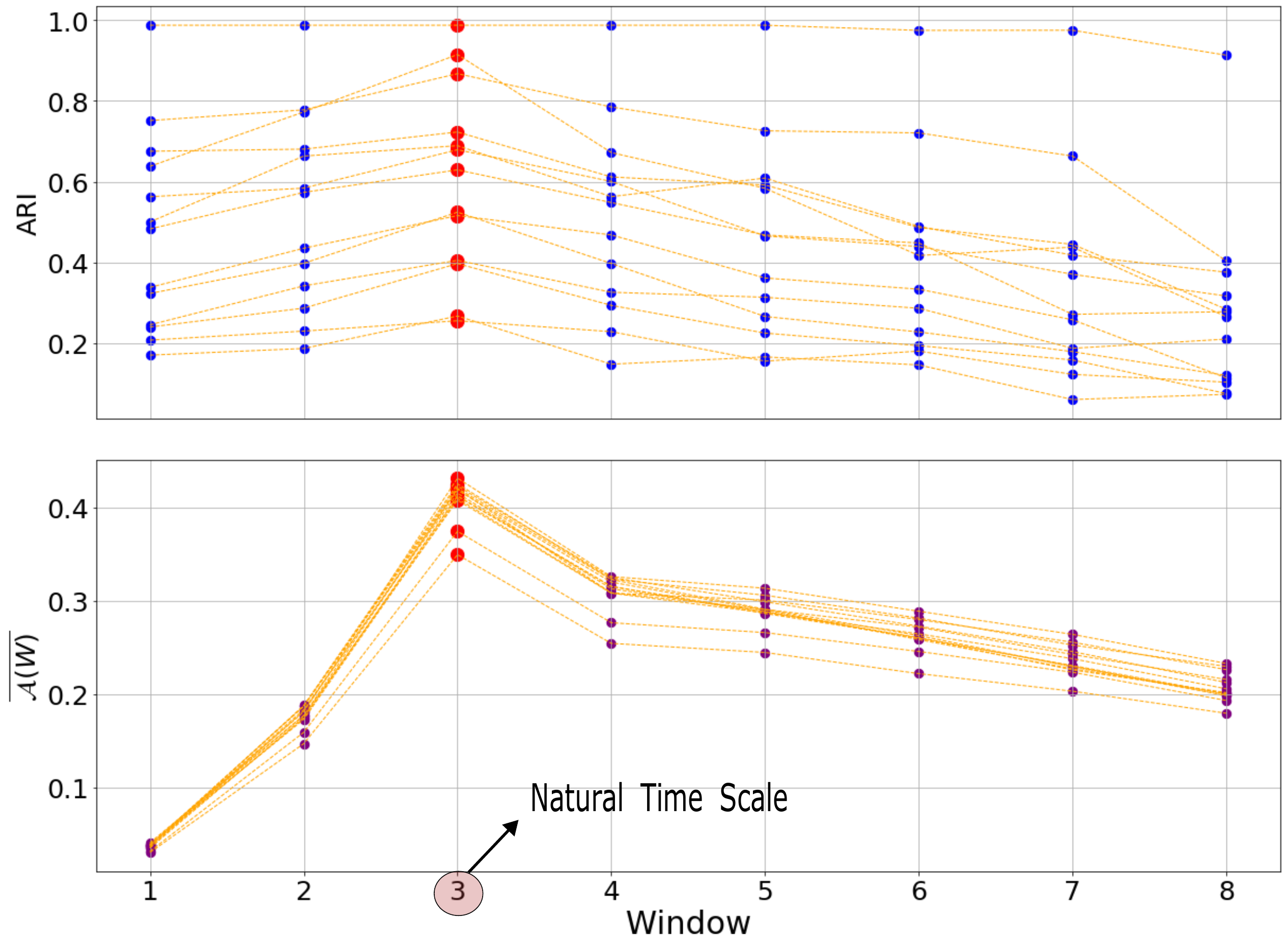}
\caption{Results from the numerical simulations described in Sec. (\ref{subsec:Numerical_Simulations_Memory})}
\label{fig: simulations_ARI_A(W)}
\end{figure}

In Fig (\ref{fig: simulations_ARI_A(W)}) are shown the results from a series of numerical simulations varying the values of \(\Delta k\) and \(\Delta h\) in a temporal network featuring 5 ground truth communities. Each point pertains to an aggregation state of the network, selecting progressively more aggregated temporal windows as the value on the x-axis increases. The two charts display the Adjusted Rand Index (ARI) value (top) for each aggregation window and the value of normalized autocovariance (bottom). Points connected by the same line represent the same network across different aggregation windows. The points highlighted in red correspond to the peak aggregation window for the variable of interest. The natural scale of aggregation is 3, which is the baseline from which all other temporal networks were generated.


\subsubsection{Real-Data results}

Below, we present data related to the application of the model aggregated over various temporal windows. This information facilitates a comparative analysis of the results obtained using modularity with memory against those obtained without memory.

If we were to use the no-memory model to identify potential structural breaks and then conduct community detection, we would find that no breakpoints are detected for any temporal window. In this regard, it's possible to compare the efficacy of the two models in explaining the data using the Akaike Information Criterion (AIC) \cite{AIC}. By doing so, considering the structural breaks identified in the memory model and those detected by the no-memory model (none), we can compare the two models by summing the AICs for each segment. The result is a clear victory for the memory model, which consistently explains the data better across all temporal windows, as indicated by lower AIC values. When comparing the results of community detection while accounting for the breaks identified by both the memory and no-memory models, we observe that the memory model almost always outperforms the no-memory model.

\begin{widetext}
\center
\begin{table}[h!]
 \begin{tabular}{||c c c c c c c c c c c c c c c c c c c c c||} 
 \hline
 Min & 1' & 2' & 3' & 4' & 5' & 6' & 7' & 8' & 9' & 10' & 11' & 12' & 13' & 14' & 15' & 16' & 17' & 18' & 19' & 20' \\ [0.5ex] 
 \hline\hline
 ARI & 0.70 & 0.80 & 0.82 & 0.85 & 0.82 & 0.84 & 0.86 & 0.86 & 0.85 & 0.87 & 0.85 & 0.87 & 0.86 & 0.87 & 0.87 &\textbf{0.90} & 0.87 & 0.88 & 0.89 & 0.89 \\ [1ex] 
 \hline
 \hline
 \end{tabular}
\caption{The table displays the average value of ARI for the different temporal aggregation windows for the modularity in eq. (\ref{temporal_modularity_2}). The average is computed by considering the value for each individual segment indicated by the changing points.}
\label{tab:modularity}
\end{table}
\end{widetext}

In Table \ref{tab:modularity}, we show  the results of modularity with memory applied for different time window. 
The ARI associated with modularity no-memory is always \textbf{0.76} because of the fact that with this model no breaks are found. 
In the memory-less model, the data fed into the algorithm is simply a temporal average of link presence. Evidently, in our scenario, this approach influences community detection in a manner independent of the temporal window chosen, further motivated by the absence of structural breaks.

This yields a value of ARI for the memory-less modularity where, only for \( W = 1' \), the outcome exceeds that of the counterpart with memory. This suggests that at shorter temporal windows, there may be memory effects not linked to membership. As noted, such effects can potentially compromise the performance of the algorithm.

\begin{figure*}[!htb] 
\begin{minipage}{0.32\textwidth}
  \centering
  \includegraphics[width=\linewidth]{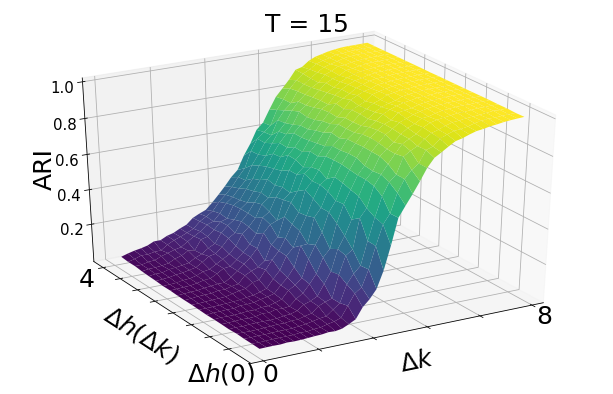}
  \caption*{(a)}
  \label{fig:no_memory_15_3d}
\end{minipage}\hfill
\begin{minipage}{0.32\textwidth}
  \centering
  \includegraphics[width=\linewidth]{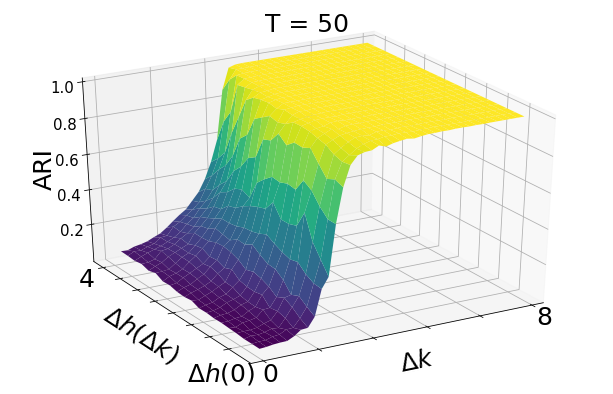}
  \caption*{(b)}
  \label{fig:no_memory_50_3d}
\end{minipage}\hfill
\begin{minipage}{0.32\textwidth}
  \centering
  \includegraphics[width=\linewidth]{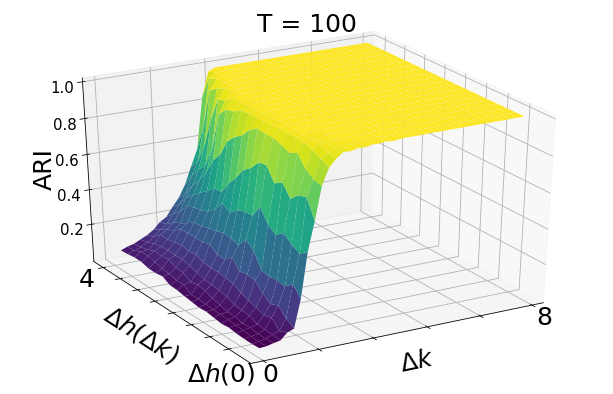}
  \caption*{(c)}
  \label{fig:no_memory_100_3d}
\end{minipage}

\begin{minipage}{0.32\textwidth}
  \centering
  \includegraphics[width=\linewidth]{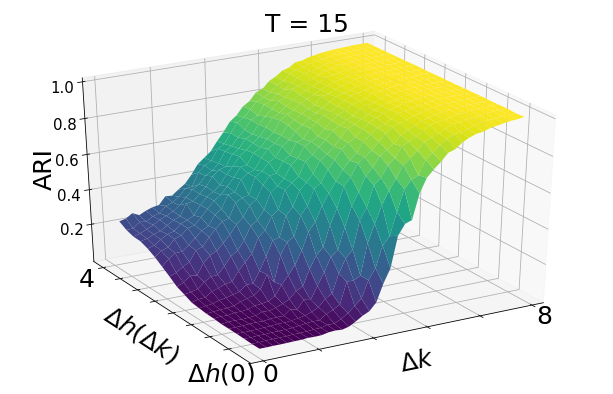}
  \caption*{(d)}
  \label{fig:memory_15_3d}
\end{minipage}\hfill
\begin{minipage}{0.32\textwidth}
  \centering
  \includegraphics[width=\linewidth]{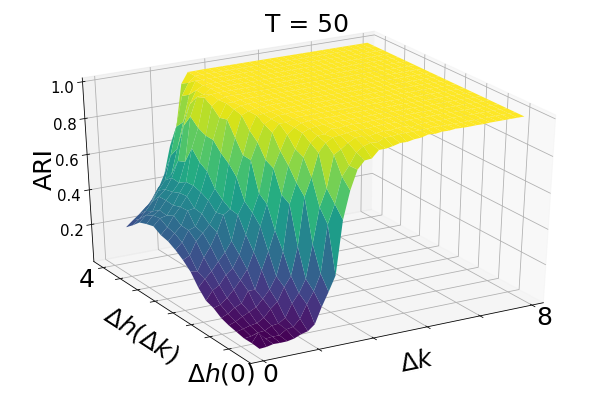}
  \caption*{(e)}
  \label{fig:memory_50_3d}
\end{minipage}\hfill
\begin{minipage}{0.32\textwidth}
  \centering
  \includegraphics[width=\linewidth]{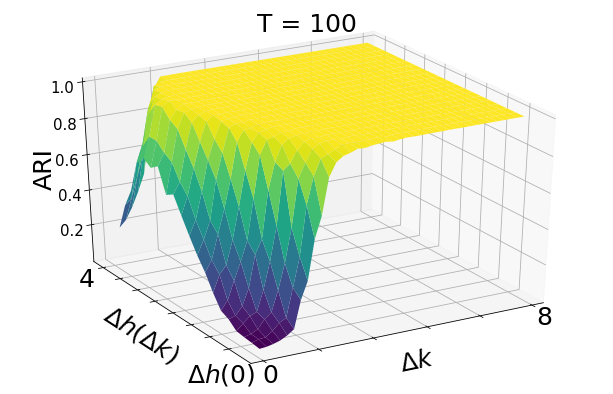}
  \caption*{(g)}
  \label{fig:memory_100_3d}
\end{minipage}

\caption{In these figures, we have a 3D representation of the average value behavior over 20 instances for \textbf{no-memory modularity} (top) and \textbf{memory modularity} (bottom) tested on graph trajectories created by varying $\Delta k$ and $\Delta h$. From left to right, we change the length of the trajectory with values T = 15, T = 50, and T = 100.}
\label{fig:combined}
\end{figure*}

\end{document}